\documentclass[sts]{pasj01}
\draft

\begin{document}
\SetRunningHead{Author(s) in page-head}{Running Head}

\Received{2017/10/06}
\Accepted{2018/04/24}{
\Published{2018/05/25}

\title{State Transitions of GRS 1739$-$278 in the 2014 Outburst}

\author{Sili WANG\altaffilmark{1},
		Nobuyuki KAWAI\altaffilmark{1},
		Megumi SHIDATSU\altaffilmark{2},
		Yutaro TACHIBANA\altaffilmark{1},
		Taketoshi YOSHII\altaffilmark{1},
		Masayuki SUDO\altaffilmark{3},
		Aya KUBOTA\altaffilmark{3}
				}

\altaffiltext{1}{Department of Physics, Tokyo Institute of Technology, 2-12-1, Ookayama, Tokyo, Japan}
\altaffiltext{2}{MAXI team, Institute of Physical and Chemical Research (RIKEN), Wako, Saitama 351-0198, Japan}
\altaffiltext{3}{Department of Electrical and Information Systems, Shibuya Institute of Technology, 307 Fukasaku, Minuma, Saitama City, Saitama 337-8570, Japan}

\email{wang.s.an@m.titech.ac.jp}

%

\KeyWords{accretion, accretion disks$-$stars: black holes$-$X-rays: binaries$-$X-rays: individual (GRS 1739$-$278) } 

\maketitle


\begin{abstract}
 We report on the X-ray spectral analysis and time evolution of GRS 1739$-$278 during its 2014 outburst based on MAXI/GSC and Swift/XRT observations. Over the course of the outburst, a transition from the low/hard state to the high/soft state and then back to the low/hard state was seen. During the high/soft state, the innermost disk temperature mildly decreased, while the  innermost radius estimated with the multi-color disk model remained constant at $\sim18\ (\frac{D}{8.5\ \mathrm{kpc}}) \ {(\frac{\cos i}{\cos 30^{\circ}})}^{-1/2}$ km, where $D$ is the source distance and $i$ is the inclination of observation. This small innermost radius of the accretion disk suggests that the central object is more likely to be a Kerr black hole rather than a Schwardzschild black hole. Applying  a relativistic disk emission model to the high/soft state spectra, a mass upper limit of $18.3\ \mathrm{M_{\odot}}$ was obtained based on the inclination limit $i<60^{\circ}$ for an assumed distance of 8.5 kpc. Using the empirical relation of the transition luminosity to the Eddington limit, the mass is constrained to $4.0-18.3\ \mathrm{M_{\odot}}$ for the same distance. The mass can be further constrained to be no larger than $9.5\ \mathrm{M_{\odot}}$ by adopting the constraints based on the fits to the NuSTAR spectra with relativistically blurred disk reflection models  (Miller et al.\ 2015).
 \end{abstract}
\begin{figure} 
\begin{center}
   \includegraphics[width=0.9\linewidth, height=0.9\textheight]{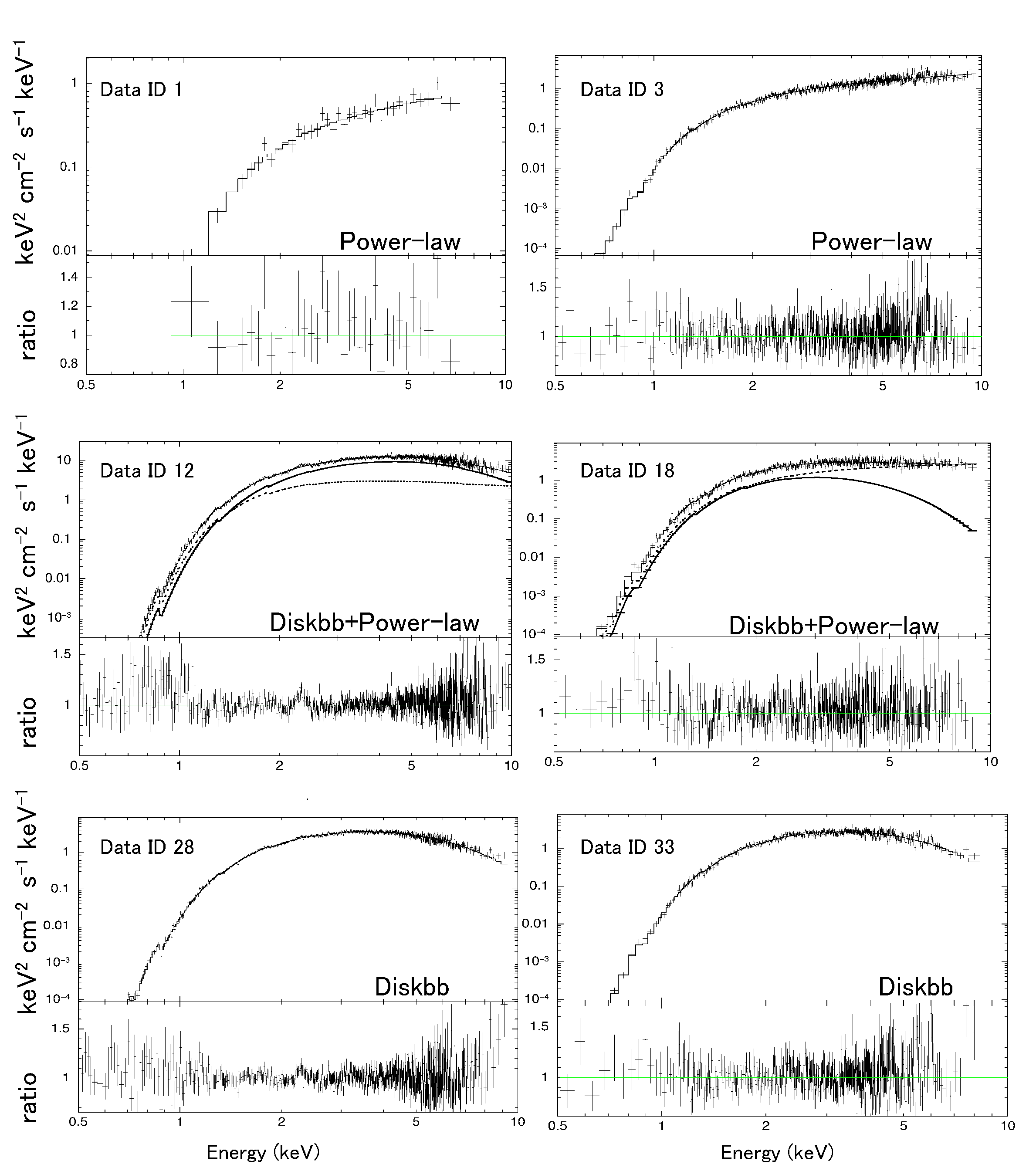} 
\end{center} 
 \caption{Examples of $\mathrm{\nu F_{\nu}}$ spectra of GRS 1739$-$278 observed with Swift/XRT shown with the best-fit models. Data ID 1 and 3 are represented by power-law models, data ID 12 and 18 represented by \texttt{diskbb+power-law} models, and data ID 28 and 33 by \texttt{diskbb} models. The dotted lines represent the power-law component while the thick solid lines represent the disk component in the spectra of ID 12 and 18.}
  \label{fig:spectra}
\end{figure}



\section{Introduction}
X-ray fluxes of black hole candidates (BHCs) at their outbursts exceed their quiescent levels by many orders of magnitude. Many BHCs exhibit state transitions, associated with changes in their X-ray fluxes during their outbursts. The presence of two different ``states" (low/hard state and high/soft state) in the X-ray emissions of the first BHC, Cyg X-1, was discovered in the early 1970s (Tananbaum et al.\ 1972). Since then, more BHCs  (e.g. GX 399$-$4, GS 2000+251; Markert et al. 1973, Tsunemi et al. 1989) have been observed to exhibit one or both of these states. Analyzing state transitions of a BHC can help us learn more about the physics of black hole accretion flows over a wide range of mass accretion rate (Remillard \& McClintock\ 2006). 
Furthermore, important characteristics about the BHC such as the black hole mass can be extracted assuming that the inner disk radius obtained in high/soft state reaches at the innermost stable orbit (Nakahira et al.\ 2012). In previous studies, light curves, hardness intensity diagrams (HID: a plot of X-ray intensity versus X-ray hardness showing evolutionary track(s); Fender et al.\ 2004), and photon spectra are frequently used to understand the nature of black hole binaries. 
Moreover, analysis of three interacting spectral components, thermal blackbody-like component, hard power-law component, and reflection component, can provide constraints on the source properties including the spin parameter and geometry. Estimates of the spin parameter were principally obtained by modeling the thermal continuum emission of the accretion disk (e.g. Zhang  et al. 1997; McClintock et al. 2014), or relativistically-broadened reflection spectrum (e.g. Fabian et al. 1989; Reynolds 2014). 
Meanwhile, a reflection spectrum reveals information about the scales of inner disk and corona. (Steiner et al.\ 2016).

GRS 1739$-$278 was first discovered in the direction of the Galactic Center with the SIGMA telescope onboard GRANAT (Paul. et al.\ 1996; Vargas et al.\ 1997). 
Its position close to the Galactic Center at $6-8.5$ was indicated by dust scattering halo in X-ray (Greiner et al.\ 1996). The new source was verified by the detection of a strong radio emission during its 1996 outburst (Hjellming et al.\ 1996). 
The distance, 8.5 kpc, is preferred according to the study of its candidate optical and infrared counterpart (Marti et al.\ 1997). 
Later, Borozdin et al.\ (1998) classified GRS 1739$-$278 as a BHC through the spectral analysis of RXTE data, in which a 5-Hz quasi periodic oscillation (QPO) was discovered (Borozdin \& Trodolyubov 2000; Wijnands et al.\ 2001). 
Following an extended quiescent period, GRS 1739$-$278 was detected in outburst with Swift Burst Alert Telescope in March 2014 (Krimm et al.\ 2014) and by INTEGRAL (Filippova et al.\ 2014). 
After this event, Miller et al.\ (2015) presented the spectral analysis of the NuSTAR observation, and gave a constraint on the innermost radius, $r_{\mathrm{in}}=5^{+3}_{-4}\ \mathrm{GM/c^2}$ as well as a spin constraint, $a=0.8\pm0.2$ during the raising part of  the ``low/hard" state.
F\"urst et al.\ (2016) also reported on a spectral analysis of the NuSTAR data at the very faint ``low/hard" state near the end of the outburst.
Mereminskiy et al.\ (2017) analyzed the latest outburst of GRS 1739$-$278 in September 2016 using INTEGRAL and Swift/XRT observations and derived a hydrogen column density $N_\mathrm{H}$ as $2.3\times10^{22}\mathrm{\ cm^{-2}}$ from spectral fitting.

In this paper we report on the spectral analysis of GRS 1739$-$278 using the Swift/XRT and MAXI/GSC data and describe the time evolution of its X-ray properties during the 2014 outburst. 
We show a mass constraint of the central object based on the spectral fittings during the high/soft state. 
We then compare the calculated bolometric luminosities to the Eddington Luminosity at various phases of the outburst.


\section{Observations and Data Reduction}

In the present analysis, we employed the data taken by Swift/XRT (Burrows et al.\ 2005) and MAXI/GSC (Mihara et al.\ 2011).

The Swift/XRT observations of the 2014 outburst of GRS 1739$-$278 were carried out during the period from 20 March 2014 (UTC) to 02 November 2014 (UTC). Although there are another episode of a small outburst after June 2015 covered by Swift/XRT, the flux was too low to study with MAXI.
We therefore limit the scope of this work to the prominent outburst episode in 2014 that exhibited clear state transitions.
Form March 2014 to November 2014, the Swift /XRT carried out pointed observations of GRS 1739-278 in the Window Timing (WT) mode. 
We extracted 42 segments of data and generated the light curves for these pointed observations in the energy bands of $2-4$ keV and $4-10$ keV using the Swift software version 3.7 from the HEASARC archive\footnote[1]{http://www.swift.ac.uk/user\_objects/}. 
For the same data segments, we constructed energy spectra using grade-zero events. 
Table \ref{tab:longtab} gives the time period for each data segment in MJD and the exposure time. In each spectrum, we combined adjacent energy bins until they satisfied a 30 photon counts threshold before we fit them with XSPEC 12.8.1. 
We adopted the energy rage of $0.5-10$ keV for all the Swift/XRT spectra .

The MAXI mission (Matsuoka et al. 2009) started on 7 Augest 2009 (UTC) (Mihara et al.\ 2011; Tsunemi et al.\ 2010), but no significant flux from GRS 1739$-$278 was recorded in the public MAXI light curve\footnote[2]{http://maxi.riken.jp/top/lc.html} until the 2014 outburst started on March 2014. 
We generated the MAXI/GSC photon event data using the on-demand software version 2.0 \footnote[3]{http://maxi.riken.jp/mxondem/} with 7-day binning. We carefully extracted photon events for the source from a circular region with a $1.^{\circ}0 $ radius and for the background with a $3.^{\circ}0$ radius, excluding the regions around nearby sources (SAX\ J1747.0-2853, GX\ 3+1, 1A\ 1742-294, and Terzan\ 5) with $1.^{\circ}$6 radii. Light curves in energy bands of $2-4$ keV and $4-10$ keV together with the corresponding HID were constructed for the outburst period from MJD 56700 to MJD 56996. We did not employ the data after MJD 56996, since the object was so faint and even slight contamination from nearby bright sources could cause significant uncertainties in flux. Some of these MAXI/GSC data were also used to create a energy spectrum in the $2-20$ keV range towards the end of the outburst from MJD 56966 to MJD 56994 when the Swift/XRT data were not available.
\section{Results and Analysis}

\subsection{Light Curves and HIDs}

In figure \ref{fig:lc} we show the X-ray light curves of GRS 1739$-$278 measured with Swift/XRT and MAXI/GSC during the outburst in $2-4$ keV energy band and $4-10$ keV energy band, together with the hardness ratio as a function of observation time in MJD, where the time of each data point is the midpoint of the start and end dates in table \ref{tab:longtab}. The Swift/XRT coverage of the outburst is shorter than the MAXI/GSC observation, but both observations reveal more variable photon counts in the first half of the outburst with an obvious flux-peak around MJD 56781, followed by a gradual, less variable decay in the second half. Both Swift/XRT and MAXI/GSC hardness ratios generally exhibit steady decrease from the beginning of the outburst until the end of Swift/XRT observation at MJD 56963. After then, MAXI/GSC hardness ratio shows the indication of returning to a hard state.

Figure \ref{fig:hid} shows the HIDs during the outburst. The Swift/XRT HID forms an upper part of a counter-clockwise circle showing  state transitions from the low/hard state to the bright intermediate state then to the high/soft state. In comparison, the MAXI/GSC reveals a complete counter-clockwise circle, which in addition shows a state transition from the high/soft state to the low/hard state at the end. The data points above the solid line in MAXI/GSC HID are those taken in the same time interval as the Swift/XRT observations. During the bight phase (MJD 56750$-$ MJD 56870), the intensity varied greatly despite of relatively small change of hardness ratio in both HIDs. As we see in the following spectral analysis, the large intensity variation is primarily caused by the disk component.

\subsection{Spectral Analysis}

We fitted 42 Swift/XRT spectra and a MAXI/GSC spectrum in XSPEC 12.8.1 with four models: (1) a power-law model; (2) a \textit{combined} model consisting of a multi-color disk (MCD; \texttt{diskbb} in XSPEC; Mitsuda et al.\ 1984) and a power-law component representing the hard tail; (3) a \textit{convolved} model consisting of a MCD model modified with Comptonization  (\texttt{simpl} in XSPEC; Steiner et al.\ 2009); (4) a MCD model. Table \ref{tab:longtab} lists the results of the best-fit model(s), which we chose based on the smallest $\chi^2_{\mathrm{\nu}}$ as well as the acceptable ranges for the hydrogen column density ($1.5 < N_{\mathrm{H}} < 2.5$. Greiner et al. 1996; Mereminskiy et al.\ 2017), the innermost disk temperature ($0.1\ \mathrm{keV} < T_{\mathrm{in}} < 3.0\ \mathrm{keV}$) and  the scattering fraction ($0<\mathrm{FracStr}\leqq1$). More than one models may be listed if the difference of reduced $\chi^2_{\mathrm{\nu}}\leq 0.02$, which corresponds to $\Delta \chi^2_{\mathrm{\nu}}\leq 8-14$ for dof (degrees of freedom): $451-753$.} Errors are quoted at the statistical 90\% confidence limits for a single parameter. Figure \ref{fig:spectra} lists some examples of Swift/XRT spectra with best-fit models. We found that outburst can be divided into four spectral states based on the requirement of spectral model as described below.

(1) The low/hard state (MJD 56736$-$MJD 56746, data IDs 1$-$4): the power-law model fits the spectra well with smaller photon index values ($\Gamma\sim1.4$) during this period. This power-law component may correspond to the ``lamp post" corona discussed by Miller et al.\ (2015) found in the NuSTAR data  around MJD 56743 (near our 4th data segement). 

(2) The intermediate state (MJD 56746$-$ MJD 56870, data IDs 5-26): complex models containing a MCD component generally show better fits during this period. We find that both the power-law model and the comptonized model, \texttt{simpl}, can represent the hard tail equally well in all these spectra. We also note that the photon index of the power-law component increased to $1.8-2.0$ as the disk component emerged, while the \texttt{simpl} model gave even larger photon index, $\Gamma>2.2$. At the brightest phase (MJD 56781, data ID 12) the \texttt{simpl} model gave a very large photon index of 4.1 while the photon index of the power-law component in the \textit{combined} model is consistent with those of other spectra. At this point we observed the maximum flux of the outburst, $F_{\mathrm{max}}=(5.75\pm0.03) \times 10^{-8}\ \mathrm{erg\ cm^{-2}\ s^{-1}}$, based on \texttt{diskbb+power-law} model. Use of \texttt{simpl} model results in a similar value. When the object comes to the end of the intermediate state (MJD 56806$-$MJD 56870, data IDs19 $-$ 26), the spectra do require a non-thermal component, but sometimes the non-thermal component is too weak to constrain its photon index. Therefore, we fixed the photon index to 2.0 for some spectra. Fixing photon index does not influence the spectral parameters of the disk component and $\chi^2_{\mathrm{\nu}}$. At the end of the intermediate state, we find that the power-law normalization is decreasing in a \textit{combined} model of \texttt{diskbb+power-law}, so does the fraction of scattering in a \textit{convolved} model of MCD with the comptonization (\texttt{simpl}). This suggests the disk component is becoming dominant.

\begin{figure} 
\begin{center}
\includegraphics[width=\linewidth]{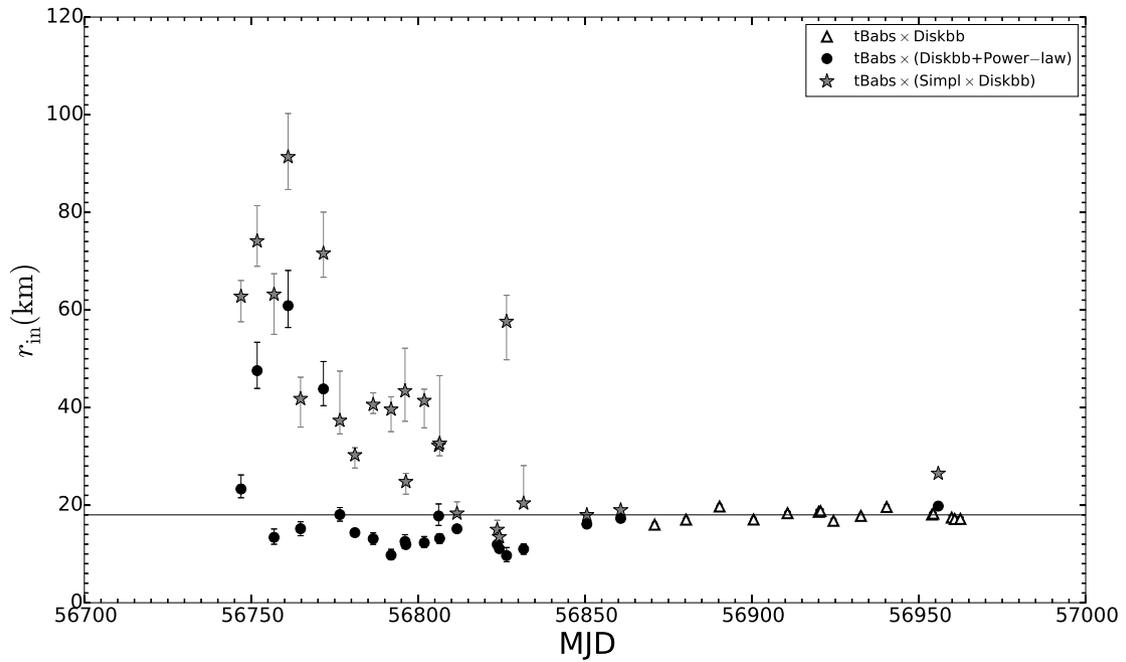}
\end{center}
 \caption{Innermost radius $r_{\mathrm{in}}$ of GRS 1739$-$278 during the intermediate state and the high/soft state using Swift/XRT spectra assuming that distance is 8.5 kpc and the inclination angle $i=30^{\circ}$.Solid horizontal line shows the weighted average of the innermost disk radius at the high/soft state, $r_{\mathrm{in}}=18$ km. The errors of are calculated based on the 90\% errors of disk normlization in spectral fittings. When two models give good fits, we present $r_{\mathrm{in}}$ for both models (see table \ref{tab:longtab}).}
  \label{fig:rin}
\end{figure}

\begin{figure} 
\begin{center}
\includegraphics[width=\linewidth]{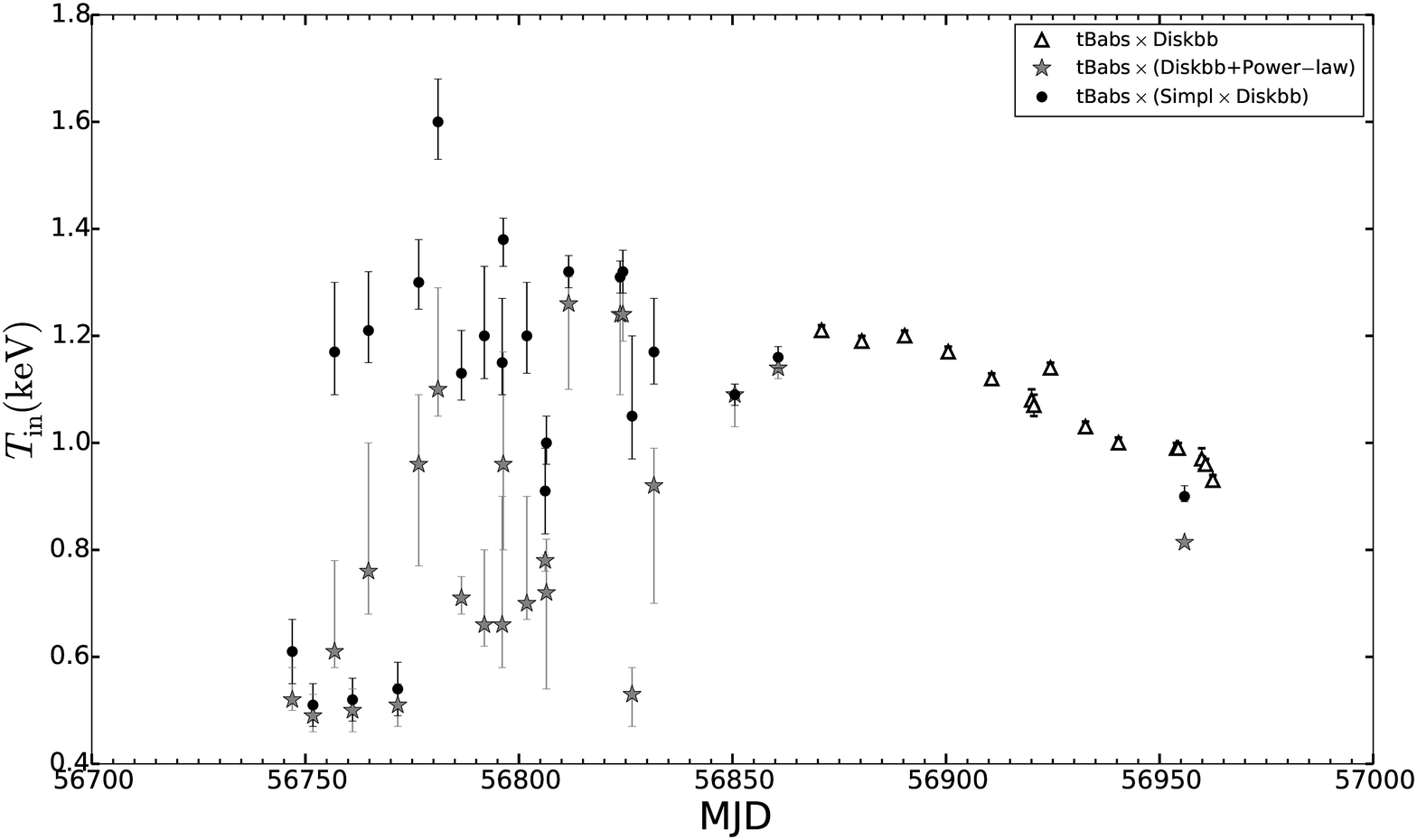}  
\end{center}
 \caption{Innermost temperature $T_{\mathrm{in}}$ of accretion disk of GRS 1739$-$278 during the intermediate state and the high/soft state obtained using Swift/XRT spectra.The errors are quoted at the statistical $90\%$ confidence limits. When two models give good fits, we present $T_{\mathrm{in}}$ for both models (see table \ref{tab:longtab}).}
  \label{fig:tin}
\end{figure}

(3) The high/soft state (MJD 56870$-$MJD 56966, data IDs 27$-$42): the non-thermal component was found to be not necessary during this period, although addition of a non-thermal component slightly improves the fitting for some spectra. The two complex models give very similar parameters for the thermal component as those given by MCD-only model, while the power-law normalization or the fraction of scattering dropped to zero except for the spectrum whose data ID is 39. We consider GRS 1739$-$278 was mostly in the high/soft state during this period. In the high/soft state, the inner disk temperature roughly shows a decreasing tendency (see figure \ref{fig:tin}) while the disk normalization stays stable (see figure \ref{fig:rin}). This indicates that the mass accretion rate decreased in this period.

(4) The transition back to the low/hard state ( MJD 56966$-$MJD 56994, data ID 43): for this data segment, we employed a MAXI/GSC spectrum due to the lack of the Swift data. The object was faint at this time and the MAXI/GSC spectrum obtained was not as ideal as Swift/XRT's in terms of statistics. Nevertheless, we find the complex model containing non-thermal component give significantly better fit than a MCD-only model. 
Here, the hydrogen column density ($N_{\mathrm{H}}$) was fixed to $2.0\times10^{22}\ \mathrm{cm^{-2}}$ considering the fitting results of Swift/XRT spectra and the results from previous researches (Greiner et al.\ 1996;  Mereminskiy et al.\ 2017). The appearance of a non-thermal component and the decrease of the inner disk temperature in spectral fittings, as well as the increase of hardness ratio in MAXI/GSC data, indicate that the object was returning to the low/hard state.

\begin{figure} 
\begin{center}
\includegraphics[width=\linewidth]{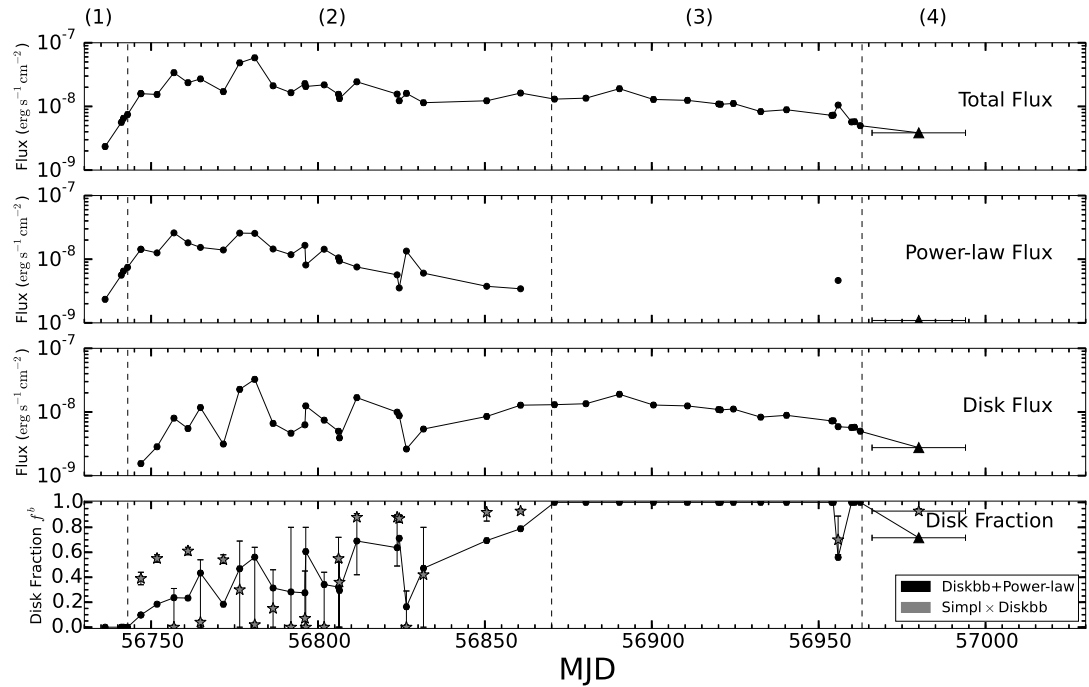}  
\end{center}
 \caption{Total flux, power-law flux, disk flux and disk fraction $f^b$ of GRS 1739$-$278 in 0.5--100.0 keV band during  the 2014 outburst. Four regions divided by dashed lines are: (1) the low/hard state (MJD 56736$-$MJD 56746); (2) the intermediate state from the low/hard state to the high/soft state (MJD 56746$-$MJD 56870); (3) the high/soft state (MJD 56870$-$MJD 56966); (4) the transition from the high/soft state to the low/hard state (MJD 56966$-$MJD 56994). Disk fraction $f^b$ is equal to zero when there is no thermal component or the fraction of scattering in \texttt{simpl} model is unity, while $f^b$ equals to unity when only the thermal component is dominant and is represented by MCD model. Only the points falling in the transition period are derived from the MAXI observation, while the others are from Swift/XRT.}
  \label{fig:diskf}
\end{figure}

To learn more about the state transitions of GRS 1739$-$278 during the outburst, we calculated the disk flux, non-thermal flux modelled by power-law, and the total flux (all unabsorbed) in $0.5-10$ keV ($2-20$ keV for MAXI/GSC data) based on the best-fit models. Figure \ref{fig:diskf} shows the variations of the disk fraction $f^b$ (the ratio of disk flux to total flux) during the outburst (For the \textit{convolved} model, \texttt{simpl$\times$diskbb}, we use 1$-$FracStr to represent the disk fraction, where the FracStr is a XSPEC model parameter: the fraction of scattering, representing the comptonization contribution in the \textit{convolved} model: \texttt{simpl$\times$diskbb}). From figure \ref{fig:diskf}, we can see that the disk fraction started from zero at the initial low/hard state then varied between zero and unity in the intermediate state, and stayed mostly at unity during the high/soft state, finally decreased. The \textit{combined} model and the \textit{convolved} model give different values of disk fraction while \textit{convolved} model has larger uncertainties. Nevertheless, the evolution of disk fraction discussed above the evolution of disk fraction discussed above, which illustrates the changes of emission component, is more or less consistent in the two models
\section{Discussion}

\subsection{State Transitions}

X-ray states have been noticed in similar spectral transitions exhibited by Cyg X-1 and A0620-00 (Coe et al, 1976). Although Cyg X-1 is not a good choice as a prototype because its soft state is not consistent with a thermal interpretation (Zhang et al.\ 1997), follow-on researches showed that states could be distinguished through differences in their photon spectral indices, luminosities, and power density spectra. Remillard \& McCclintock (2006) found that the high/soft state has a disk fraction $f^b>75\%$, while the low/hard state has a disk fraction $f^b<20\%$ when using the \textit{combined} model, \texttt{diskbb$+$power-law}. In their work, they defined disk fraction as the ratio of disk flux to the total flux (both unabsorbed) at $2-20$ keV, which is somewhat different from our energy range. Here we defined the high/soft state to be when the spectrum can be modeled only by the disk component and when the inclusion of the non-thermal component does not improve the fittings. The opposite situation models the low/hard state. As we described in section 3, we divided the outburst into four parts: (1) The low/hard state (MJD 56736$-$MJD 56746, when only the non-thermal component is required); (2) The intermediate state during the transition from the low/hard state to the high/soft state (MJD 56746$-$MJD 56870, when both thermal and non-thermal components are required); (3) The high/soft state (MJD 56870$-$MJD 56966, when only thermal component is required); and (4) The transition from the high/soft state to the low/hard state (MJD 56966$-$MJD 56994, when both thermal and non-thermal components are required again).

In the low/hard state of GRS 1739$-$278, the photon index varied around 1.40$-$1.43, consistent with the definitions of the low/hard state given by Remillard \& McClintock (2006): $1.4<\Gamma<2.1$. Also the innermost disk temperature in the high/soft state falls into the typical range, 0.75 $-$ 1.5 keV. During the intermediate state, situations are quite complex. According to the results of the \textit{combined} model, \texttt{diskbb$+$power-law}, the disk fraction generally falls within the range of $0.2-0.8$ during the intermediate state. On the other hand, the disk fraction shows a different performance based on the \textit{convolved} model, \texttt{simpl$\times$diskbb}. Its results suggest relatively stronger contribution to X-ray emissions from Comptonization corresponding to a huge corona that enveloped the accretion disk. While the timing information, like QPO, could be a good tracer for the different X-ray states (Remillard \& McClintock.\ 2006; Sobczak et al.\ 2000), it is beyond the scope of the present work and left to the future study.

\subsection{Innermost Radius and Temperature of Accretion Disk}

Figure \ref{fig:rin} shows the variation of the innermost disk radius, $r_{\mathrm{in}}$, that is given according to the relationship between the disk normalization and innermost radius (Mitsuda et al.\ 1984; Makishima et al.\ 1986):
\begin{equation}
r_{\mathrm{in}} (\mathrm{km}) = Norm^{1/2} \ (\frac{D}{8.5\ \mathrm{kpc}}) \ {(\frac{\cos i}{\cos 30^{\circ}})}^{-1/2},
\end{equation}
where $Norm$ is the disk normalization, $D$ is the distance and $i$ is the inclination angle. During the intermediate state, the \textit{convolved} model:\texttt{simpl$\times$diskbb} and the \textit{combined} model: \texttt{diskbb$+$power-law} reveal different behaviors of $r_{\mathrm{in}}$. From a broad view of time, based on the \textit{convolved} model we find that the accretion disk gradually extended inward during the intermediate state (MJD 56746$-$MJD56861) and reached to the innermost radius that remained almost constant in the high/soft state. On the other hand, in the \textit{combined} model, the accretion disk shrank quickly to the same innermost radius at the beginning of the intermediate state (MJD 56746$-$ MJD 56777) then varied with small fluctuations. The innermost temperature of the accretion disk, $T_{\mathrm{in}}$, also showed different evolutions between the \textit{convolved} model and the \textit{combined} model in the intermediate state. 
Generally, $T_{\mathrm{in}}$ increased with a wide fluctuation during this period based on the \textit{convolved} model. From the results of the \textit{combined} model, $T_{\mathrm{in}}$ also increased with a wide fluctuation at the very beginning of the intermediate state but showed a decrease tendency after the brightest phase (MJD 56781, Data ID: 12). Although the \textit{convolved} model is physically understandable, the data analysis shows that the \textit{combined} model is more consistent with the results in the high/soft state when the accretion disk was dominant (i.e. the hydrogen column density and the innermost disk radius). This suggests the existence of a separate non-thermal component rather than the comptonized disk emission during the intermediate state. 
Although the two complex models indicate different evolutions of the accretion disk during the intermediate state, the parameters of disk component converge to remarkably similar values in both models at the end of the intermediate state and were smoothly connected to those in the high/soft state. 

\subsection{Mass Constraint of Central Object}

\begin{figure} 
\begin{center}
  \includegraphics[width=\linewidth]{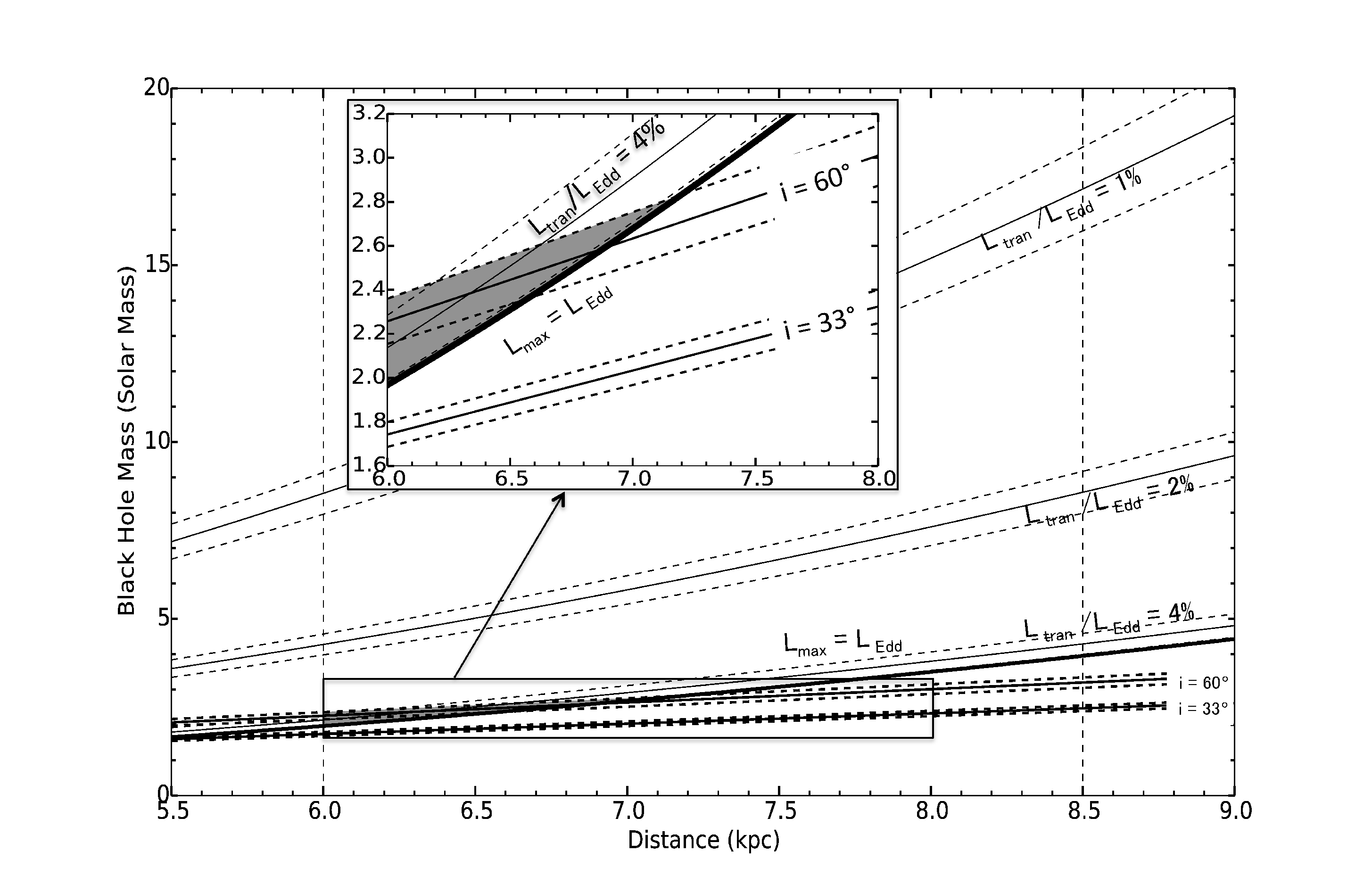}
\end{center}
  \caption{Constraints on the mass and the distance of GRS 1739$-$278 assuming a non-spinning black hole. The shadowed region in the inset shows the allowed mass region constrained by the maximum luminosity and the innermost radius from Swift/XRT spectra assuming the inclination range of $33^{\circ}-60^{\circ}$. Each solid curve labeled with $L_{\mathrm{tran}}/L_{\mathrm{Edd}}$ indicates the constraint on black hole mass for different Eddington ratios (1\%, 2\%, 4\%) of the transition luminosity $L_{\mathrm{tran}}$. The 90\% confidence ranges are shown with accompanying dashed curves. Thick solid curve labeled with $L_{\mathrm{max}}=L_{\mathrm{Edd}}$ indicates the lower limit of black hole mass set by the maximum luminosity not exceeding the Eddington limit.}

  \label{fig:dbb}
\end{figure}

When GRS 1739$-$278 was in the high/soft state (MJD 56780$-$MJD 56963), the innermost radius $r_{\mathrm{in}}$  remained almost constant at $18.00\pm1.11\ (\frac{D}{8.5\ \mathrm{kpc}}) \ {(\frac{\cos i}{\cos 30^{\circ}})}^{-1/2}$ km (weighted average and standard deviation). Furthermore, this constancy allows us to identify $r_{\mathrm{in}}$ as the inner stable circular orbit (ISCO) in the high/soft state. 
 We note that $r_{\mathrm{in}}$ is an ``apparent" innermost radius, and the more ``realistic" innermost radius $R_{\mathrm{in}}$ should be estimated as $R_{\mathrm{in}}=\xi{\kappa}^2r_{\mathrm{in}}$, where the spectral hardening factor, $\kappa$, is 1.7 (Shimura \& Takahara.\ 1995) and correction factor for the boundary condition, $\xi$, is 0.412 (Kubota et al.\ 1998). 
Considering the color correction factor is found to be consistent with the canonical value ($f_{\mathrm{col}}\sim1.7$) for the majority of CCD (Swift/XRT) observations (Reynolds \& Miller.\ 2013), our correction with $\kappa=1.7$ should be representative for the discussions of innermost disk radius and black hole mass during high/soft state when the accretion disk dominates. When the central object is assumed to be a non-spinning black hole, $R_{\mathrm{ISCO}}$ should be equal to 6 $R_{\mathrm{g}}$ (where $R_{\mathrm{g}} =\frac{GM^2}{c^2}$), yielding a  black hole mass of $2.46\pm0.07 \ \mathrm{M_{\odot}}$ with the assumption that distance is 8.5 kpc (Marti et al.\ 1997) and the inclination angle is $i=33^{\circ}$ (Miller et al.\ 2015). This mass is smaller than those found in luminous stellar-mass BHCs and the maximum luminosity at brightest phase would exceed the Eddington luminosity. The shaded region in figure \ref{fig:dbb} shows the constraint on the mass based on the conditions that $R_{\mathrm{in}}=R_{\mathrm{ISCO}}$ and $L_{\mathrm{max}}\leq L_{\mathrm{Edd}}$, where $R_{\mathrm{in}}$ is calculated for the possible ranges of the inclination $33^{\circ}\leq i \leq 60^{\circ}$ and the  distance $6-8.5$kpc  (Dennerl \& Greiner, 1996). Even if we consider the extreme case that $L_{\mathrm{max}}=L_{\mathrm{Edd}}$ and $i=60^{\circ}$, the black hole mass should be smaller than $2.85\ \mathrm{M_{\odot}}$, which is also much smaller than the mean black hole mass of the 12 black hole transients with firm mass measurements (cf. Corral-Santana et al.\ 2016). Here the maximum possible inclination of $60^{\circ}$ is derived from the lack of dip in the light curves ( Frank et al.\ 1987). We conclude that GRS 1739$-$278 is not likely to be a Schwarzschild black hole.

\begin{figure} 
\begin{center}
  \includegraphics[width=\linewidth]{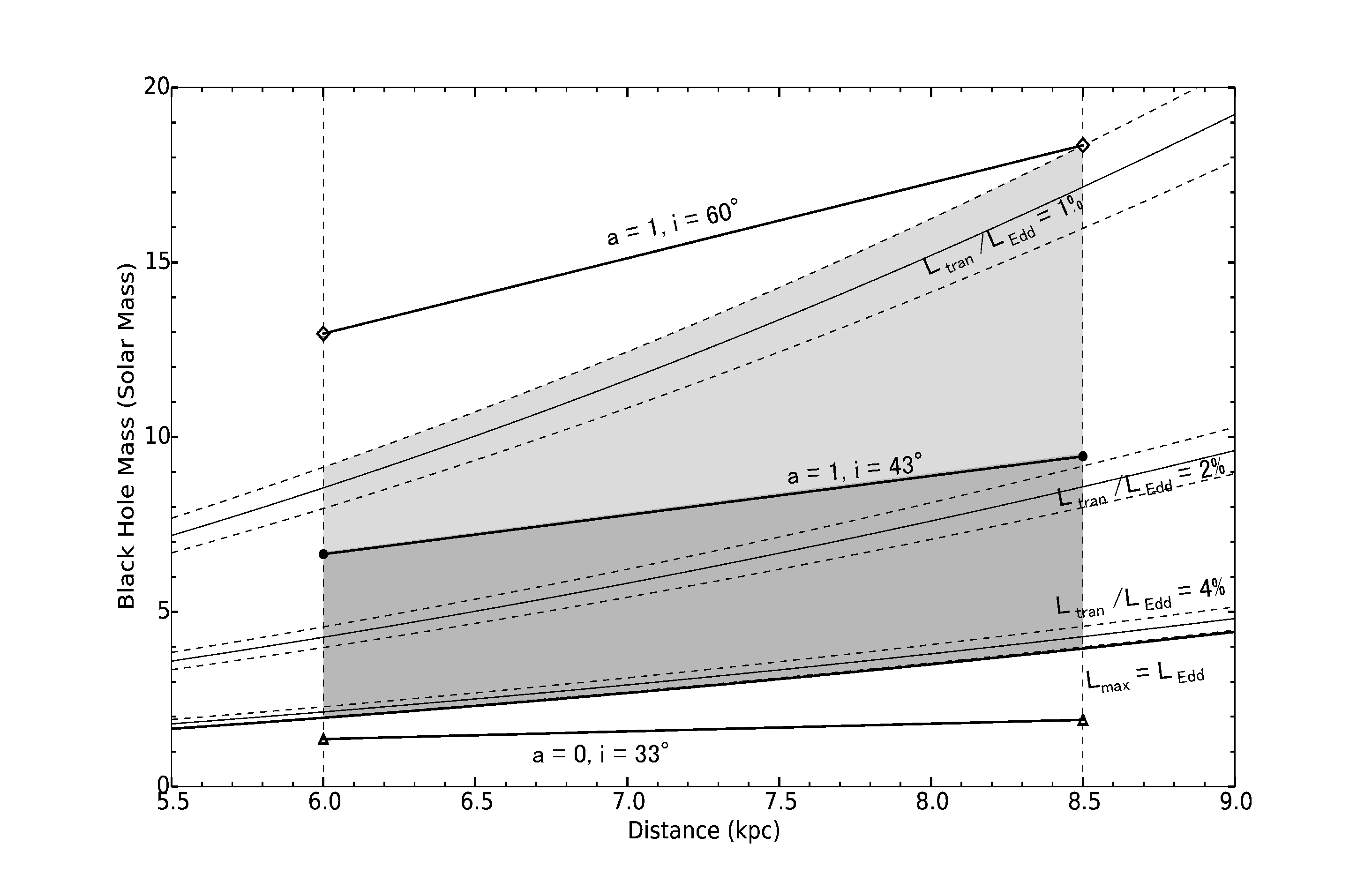}
\end{center}
 \caption{Constraints on the mass and the distance of GRS 1739$-$278 based on \texttt{kerrbb} model. The shadowed region shows the black hole mass range satisfying the conditions on observational luminosity that $1\%L_{\mathrm{Edd}}<L_{\mathrm{tran}}<4\%L_{\mathrm{Edd}}$ and $L_{\mathrm{max}}\leq L_{\mathrm{Edd}}$, while the deeper-colored part shows the range in combination with Miller et al's (2015) results of spin and inclination. The solid lines marked with symbols represent the weighted average of the spectral fitting results of 12 spectra during the high/soft state based on \texttt{kerrbb} model with the spin parameters and the inclinations shown in table \ref{tab:appen1}. Each solid curve labeled with $L_{\mathrm{tran}}/L_{\mathrm{Edd}}$ indicates the constraint on black hole mass for different Eddington ratios (1\%, 2\%, 4\%) of the transition luminosity $L_{\mathrm{tran}}$. The 90\% confidence ranges are shown with accompanying dashed curves. Thick solid curve indicates the lower limit of black hole mass set by the maximum luminosity not exceeding the Eddington limit.}
  \label{fig:kbb}
\end{figure}

Considering the suggested spin value $0.8\pm0.2$ (Miller et al.\ 2015), we may obtain a more reasonable constraint by assuming GRS 1739-278 is a Kerr black hole. To validate the result, we tested 12 spectra in the hight/soft state with a multi-temperature blackbody model for a thin accretion disk around a Kerr black hole (\texttt{kerrbb} in XSPEC; Li et al.\ 2005).
For each spectral fitting, we fixed the ratio of eta (ratio of the disk power produced by a torque at the disk inner boundary to the disk power arising from accretion) to zero, the normalization to unity, the inclination within $33^{\circ}\leq i \leq 60^{\circ}$, and the distance within a range 6$-$8.5 kpc, respectively. 
We let the hydrogen column density, the spin value (range: $0-1$), the black hole mass, and the mass accretion ratio variable, then keep the other parameters at their default values (spectral hardening factor 1.7, rflag 1.0, and  lflag 0). The Swift/XRT spectra we employed failed to constrain spin parameter in \texttt{kerrbb} model. 
As the estimated black hole mass is a monotonically increasing function of the spin parameter and inclination, we obtained the upper limit of possible black hole mass with the extreme condition that $i =60^{\circ}$ and $a=1$. The solid line labeled with $i =60^{\circ}$ in figure \ref{fig:kbb} shows the weighted averages of the results for the 12 spectra for the possible distance of $6-8.5$ kpc. 
In combination with the condition that $L_{\mathrm{max}}\leq L_{\mathrm{Edd}}$, where $L_{\mathrm{max}}=4\pi D^2 F_{\mathrm{max}}$ and $F_{\mathrm{max}}= (5.75\pm0.03) \times 10^{-8}\ \mathrm{erg\ cm^{-2}\ s^{-1}}$, we can obtain a mass range to be $2.0-12.9\ \mathrm{M_{\odot}}$ for $D=6.0$ kpc and $4.0-18.3\ \mathrm{M_{\odot}}$ for $D=8.5$ kpc. Moreover, from the past observations of BHCs and neutron stars, Maccarone (2003) found that the state transition from the high/soft to the low/hard state occurs at $1\%-4\%$ (centered at 2\%) of the Eddington luminosity. Judging from the appearance of the non-thermal component in the MAXI spectrum (Data ID: 43) and the remarkably increase of hardness ratio in the MAXI HID, we suspect GRS 1739$-$278 went back to the low/hard state around MJD 56966$-$MJD56994. 
The transition flux can be obtained at the transition phase (MJD 56987$-$MJD 56994) shown at the extreme lower left in the MAXI HID, but we found that the photon statistics is too poor to fit the spectrum. Therefore we combined the data in MJD 56966$-$ MJD 56994 to construct the spectrum and calculate the average bolometric flux. 
The transition flux was calculated to be $F_{\mathrm{tran}} = (2.53\pm0.18) \times 10^{-9}\ \mathrm{erg\ cm^{-2}\ s^{-1}}$ by scaling the average bolometric flux by the ratio ($\sim65\%$) of the photon count rates at the transition phase to that of the average.
Using the additional condition that $0.01\;L_{\mathrm{Edd}}<L_{\mathrm{tran}}<0.04\;L_{\mathrm{Edd}}$, where $L_{\mathrm{tran}}$ is calculated by $L_{\mathrm{tran}}=4\pi D^2 F_{\mathrm{tran}}$, we can reduce the upper limit of black hole mass to be 9.1 and 18.3 $ \mathrm{M_{\odot}}$ for the distance of 6.0 and 8.5 kpc respectively. 
This mass range is shown in the shadowed region in figure \ref{fig:kbb}. 
If we further adopt the constraints on the spin and the inclination by Miller et al. (2015), we can even obtain a narrower mass range by taking the union of estimated black hole masses in three sets: (1) $a=0-1, i=33^{\circ}$ (according to \texttt{relconv\_lp\_ext $\times$ xillver} and \texttt{relxill\_lp} models); (2) $a=0.6-1, i=43^{\circ}$ (according to \texttt{relxill} model); (3) $a=0.92-0.96, i=24^{\circ}$ (according to \texttt{relcon $\times$ reflionx} model). 
The union of estimated black hole mass ranges is given by the lower limit with $a=0, i=33^{\circ}$ and the upper limit with $=1, i=43^{\circ}$. 
Here the black hole masses are derived from the weighted averages of the 12 spectral fitting results (see table \ref{tab:appen1}). 
In combination with the condition that $L_{\mathrm{max}}\leq L_{\mathrm{Edd}}$, the black hole mass can be constrained within  $2.0-6.7\ \mathrm{M_{\odot}}$ for $D=6.0$ kpc, or $4.0-9.5\ \mathrm{M_{\odot}}$ for $D=8.5$ kpc, shown as the deeper-colored part of the shadowed region in figure \ref{fig:kbb}.

In addition, we can derive the lower limit for the spin parameter using the same model, assuming that $L_{\mathrm{max}}=L_{\mathrm{Edd}}$ and $i=60^{\circ}$. 
For analysis, we chose two typical spectra (data ID: 27 and 36) which gave the smallest and the largest black hole masses in above spectral fittings, respectively.  
We then fitted them by fixing black hole masses to those given by $L_{\mathrm{max}}=L_{\mathrm{Edd}}$, and let spin parameter vary from $-$1 to 1 (see table \ref{tab:appen2}). 
For small distances close to 6 kpc, the spin value can drop to zero for the spectrum (data ID 27) with a small innermost disk radius, and the spin value can even decreased to negative for the spectrum (data ID: 36) with a large innermost disk radius. 
On the other hand, a higher spin parameter is required for a larger distance. 
The spin should be larger than 0.5 (for data ID 27) or 0.22 (for data ID 36) to ensure that $L_{\mathrm{max}}$ should not exceed $L_{\mathrm{Edd}}$ for $D=8.5$ kpc (Marti et al.\ 1997). 
This result is also consistent with our reasoning that GRS 1739-278 is likely to be a Kerr black hole.

\subsection{Luminosity}

\begin{figure} 
\begin{center}
  \includegraphics[width=\linewidth]{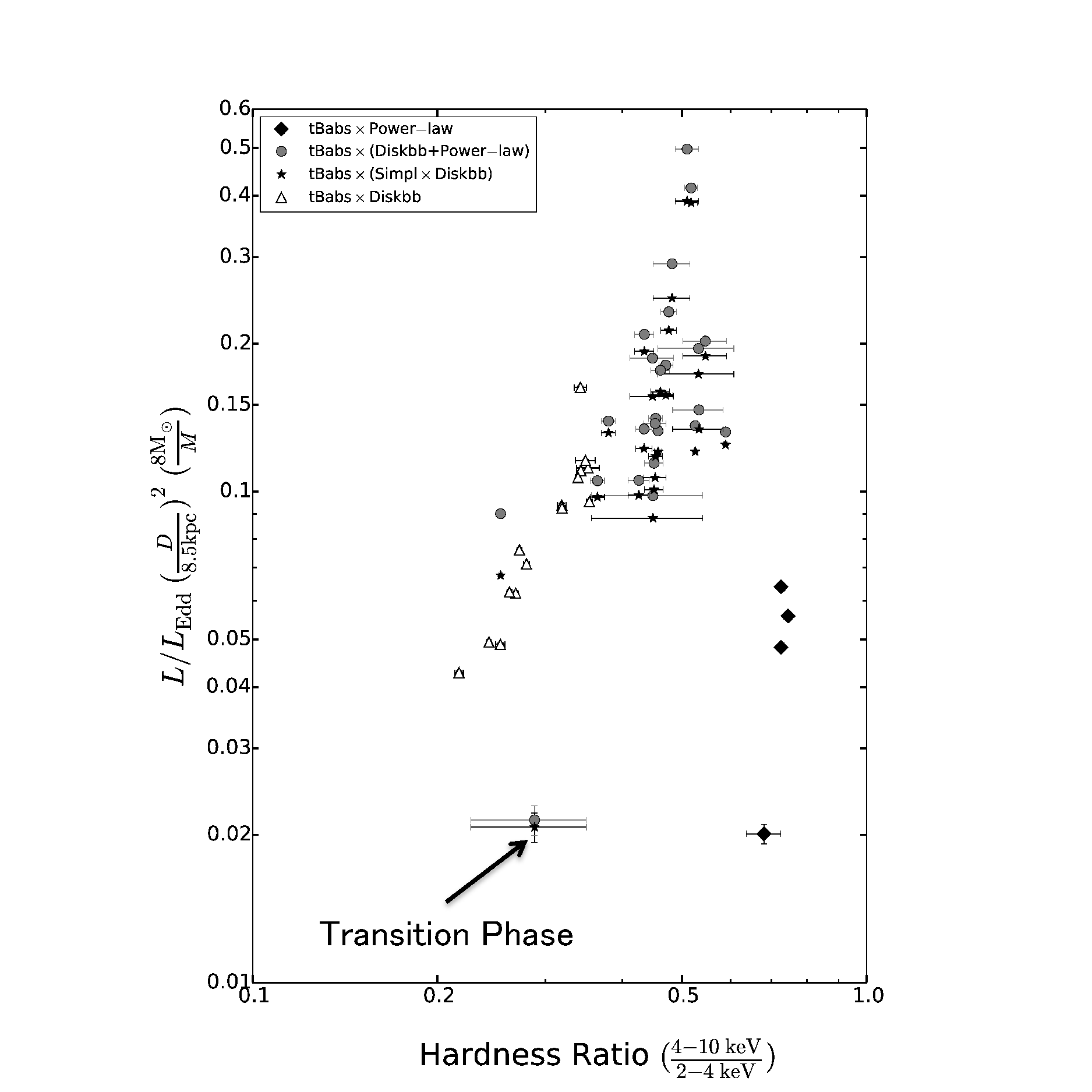}
\end{center}
 \caption{Bolometric luminosity of GRS 1739$-$278 normalized by the Eddington luminosity for a $8\ \mathrm{M_{\odot}}$ BH at a distance of 8.5 kpc. Horizontal axis shows the hardness ratio between 4--10 keV and 2--4 keV bands based on the Swift/XRT and MAXI/GSC observations. The two points locating at the extreme left bottom shows the transition luminosity based on MAXI/GSC data, while other points are based on Swift/XRT data.The errors of luminosity are calculated from the errors of flux in spectral fittings with $90\%$ confidence, but too small to be visible for most points.}
  \label{fig:lum}
\end{figure}

Figure \ref{fig:lum} is the luminosity vs. hardness ratio plot, where the bolometric luminosity is normalized by the Eddington luminosity for a 8 $\mathrm{M_{\odot}}$ black hole at a distance of 8.5 kpc. We chose 8 $\mathrm{M_{\odot}}$ for scaling because it is the mean black hole mass of Galactic black hole transients (Corral-Santana et al.\ 2016), and it gives a modal value of empirical transition luminosity, $L_{\mathrm{tran}}=0.02\;L_{\mathrm{Edd}}$. 
Here the luminosities are calculated for the $0.5 -100$ keV energy rangeby $L=4\pi D^2 F$, where the flux $F$ is determined based on the best-fit models (in Table \ref{tab:longtab}). 
We find that luminosity reached $12\%-14\% L_{\mathrm{Edd}}$ when the object came into the intermediate state (MJD 56746) and varied largely during the intermediate state. And the luminosity came back to the $12\%-14\% L_{\mathrm{Edd}}$ when the object made transition to the high/soft state (MJD 56870). The brightest phase reached around $50\% L_{\mathrm{Edd}}$.

\section{Conclusions}
Based on the Swift/XRT and MAXI/GSC observations, we analysed time evolutions of the intensities and spectra of the BHC, GRS 1739$-$278, during its 2014 outburst. We  find that the outburst can be divided into four phases based on spectral analyses: (1) The low/hard state (MJD 56736$-$MJD 56746); (2) The intermediate state during the transition from the low/hard state to the high/soft state (MJD 56746$-$MJD 56870); (3) The high/soft state (MJD 56870$-$MJD 56966); (4) The transition from the high/soft state to the low/hard state (MJD 56966$-$MJD 56994). As commonly seen in most BHCs, the innermost radius of GRS 1739$-$278 remained constant in the high/soft state. Our analysis supports that GRS 1739$-$278 is not likely to be a non-spinning black hole but rather a spinning black hole. Assuming that $a\leq1, i\leq60^{\circ}$, in combination with the previously known constraint on distance of $6-8.5$ kpc and the two conditions on the observational luminosities -- the transition from the hight/soft state to the low/hard occurs at  $1\% -4\%$ of $L_{Edd}$, the Eddington luminosity, and  the maximum luminosity should not exceed $L_{Edd}$, we constrained the mass of central object to be $2.0-18.3\ \mathrm{M_{\odot}}$ by applying \texttt{kerrbb} model to the spectra in the high/soft state. A narrower constraint was obtained when using the spin parameters and inclinations from NuSTAR's fitting results with the relativistically blurred disk reflection models (Miller et al. 2015).


In order to improve our estimate of black hole mass, we need more accurate distance, inclination, and spin parameter that may require independent observations such as imaging superluminal jets, measurement of companion radial velocity as well as better understandings of X-ray spectra.

This work is based on the data provided by Swift team and the MAXI team . Authors are grateful for the support obtained. M. S. acknowledges support by the Special Postdoctoral Researchers Program at RIKEN, and by a Great-in-Add for JSPS Young Scientist (B) 16K17672. N. K. acknowledges support by MEXT KAKENHI Grant Number JP 17H06362.


\begin{figure} 
 \begin{center}
\includegraphics[angle=90, width=0.8\linewidth, height=0.8\textheight]{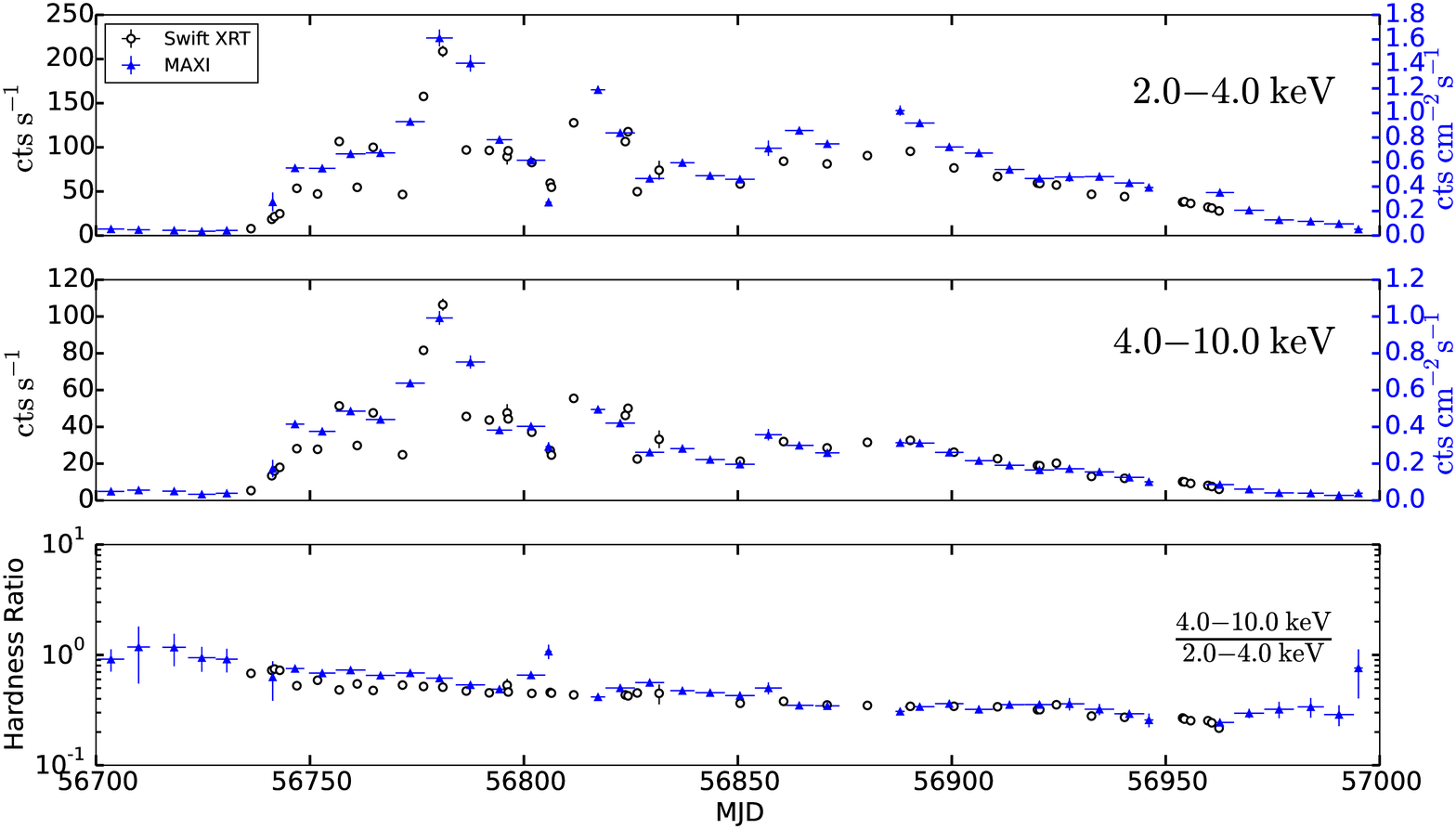}    
 \end{center}
\caption{Swift/XRT and MAXI/GSC light curves and hardness ratios of GRS 1739$-$278 in the 2014 outburst. Three panels from the top to the bottom show the light curves in 2--4 keV and 4--10 keV bands, and the hardness ratios between 4--10 keV and 2--4 keV bands. The open circles represent Swift/XRT data, while the blue triangles represent MAXI/GSC data.}\label{fig:lc}
\end{figure}

\begin{figure} 
\begin{center}
   \includegraphics[width=\linewidth]{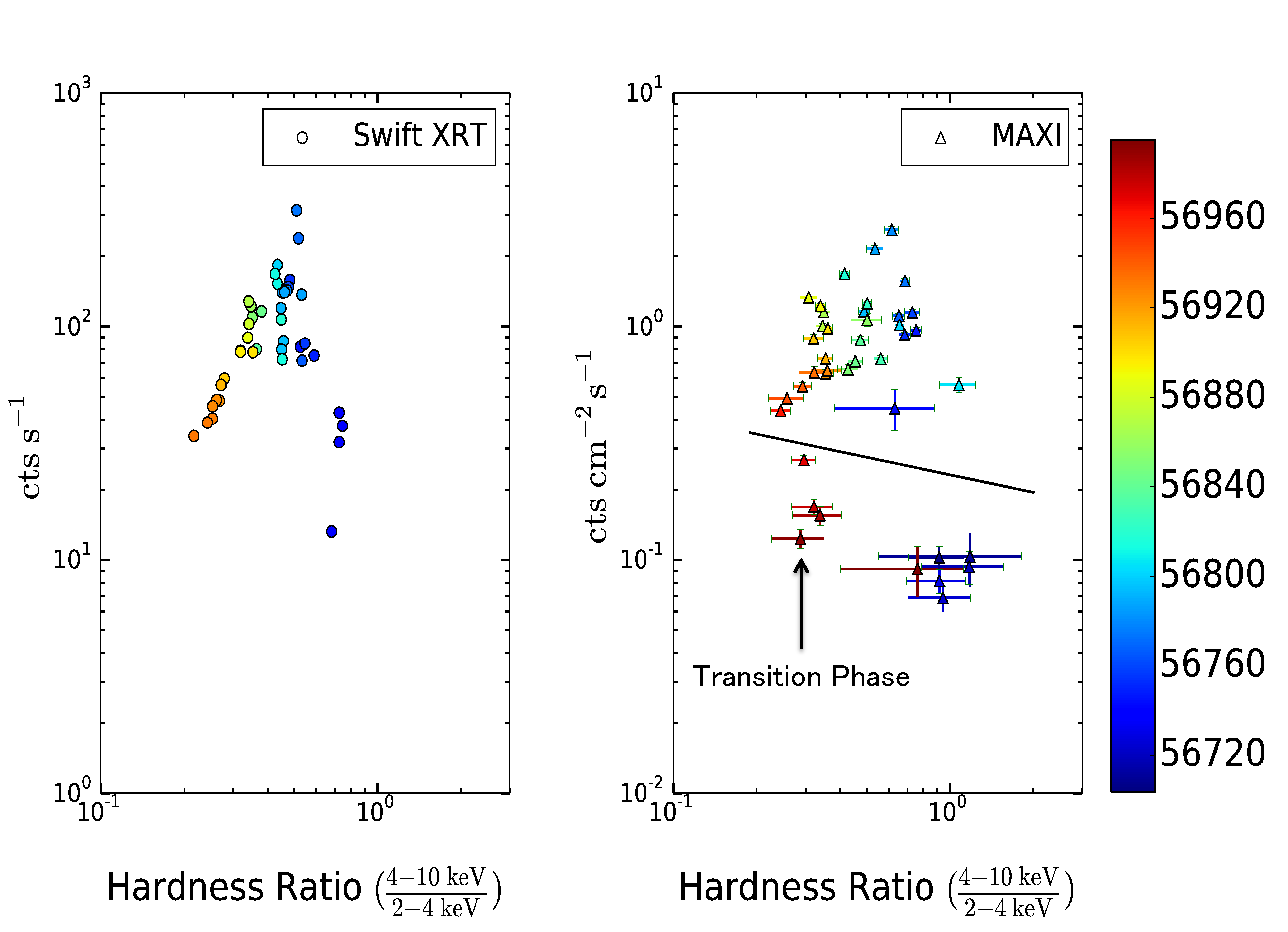} 
\end{center}
 \caption{Swift/XRT and MAXI/GSC hardness intensity diagrams (HIDs) of GRS 1739$-$278 in the 2014 outburst. Left panel shows Swift/XRT HID. Right panel shows the MAXI/GSC HID, where the data points above the solid line are those taken in the same time interval as the Swift/XRT observations. Horizontal axis represents the hardness ratio between 4--10 keV and 2--4 keV bands, and vertical axis shows the total photon counts in the 2--10 keV band. The color bar on the right presents the date encoded on the color of each data point: the change of color from blue to red represents the evolution of the outburst from the beginning to the end. The data in the transition phase from high/soft state to low/hard state at MJD 56987$-$56994 is shown with an arrow in MAXI HID.}
  \label{fig:hid}
\end{figure}

 \begin{center}
\begin {longtable}{@{\extracolsep{\fill}}ccc|cccccc}
\caption {Best-fit parameters of Swift/XRT and MAXI/GSC spectra: fitted with power-law model, comptonized MCD (\texttt{simpl$\times$ diskbb}) model, combined MCD plus power-law (\texttt{diskbb+power-law}) model, or MCD model. The rows that have " FracStr" present the results of comptonized MCD model.\label{tab:longtab}} \\
\hline
Data ID & \footnotemark[$(0)$]MJD & Exposure &\footnotemark[$(1)$]$N_\mathrm{H}$&\footnotemark[$(2)$]$\Gamma$&\footnotemark[$(3)$]$T_\mathrm{in}$&\footnotemark[$(4)$]$Norm\ of\ Diskbb$ &\footnotemark[$(5)$]FracStr 
& $\chi^2_{\nu}~(dof)$\\
&start--end& (s) &($10^{22}$~cm$^{-2}$)&&(keV)&($\mathrm{km}^2$)&&\\
\endfirsthead
\hline
Data ID & \footnotemark[$(0)$]MJD & Exposure &\footnotemark[$(1)$]$N_\mathrm{H}$&\footnotemark[$(2)$]$\Gamma$&\footnotemark[$(3)$]$T_\mathrm{in}$&\footnotemark[$(4)$]$Norm\ of\ Diskbb$ &\footnotemark[$(5)$] FracStr 
& $\chi^2_{\nu}~(dof)$\\
&start--end& (s) &($10^{22}$~cm$^{-2}$)&&(keV)&($\mathrm{km}^2$)&&\\
\hline
\endhead
\hline
\endfoot
\endlastfoot
\hline
1 & 736.213--736.214 &79.12& $1.66^{+0.36}_{-0.31}$ & $1.42^{+0.25}_{-0.24}$ & --- & --- & --- & 0.80~(33)\\
2 & 741.070--741.089 &1666.75& $1.62\pm0.04$ & $1.40\pm0.03$ & --- & --- & --- & 1.20~(631)\\
3&741.268--742.160&997.41&$1.69\pm0.05$&$1.44\pm0.04$&---&---&---&1.12~(560)\\
4&742.936--742.951&1328.03&$1.50\pm0.04$&$1.43\pm0.03$&---&---&---&1.13~(638)\\
5&746.866--747.007&2463.66&$1.70^{+0.02}_{-0.01}$&$1.93^{+0.02}_{-0.03}$&$0.52^{+0.06}_{-0.02}$&$4.7^{+1.0}_{-1.6}\times 10^3$&$0.61\pm0.05$&1.24~(753)\\
&&&$1.81\pm0.04$&$1.91\pm0.06$&$0.61\pm0.06$&
$6.5^{+3.3}_{-2.1}\times 10^2$&---&1.25~(753)\\
6&751.742--751.809&1832.34&$1.89\pm0.04$&$1.68^{+0.03}_{-0.06}$&$0.49^{+0.04}_{-0.03}$&$6.6^{+2.6}_{-1.9}\times 10^3$&$0.45\pm0.03$&1.23~(736)\\

&&&$1.96\pm0.05$&$1.62^{+0.06}_{-0.03}$&$0.51\pm0.04$&$2.7^{+1.4}_{-0.85}\times 10^3$&---&1.23~(736)\\

7&756.809--756.873&1811.94&$1.78^{+0.03}_{-0.04}$&$2.42^{+0.05}_{-0.10}$&$0.61^{+0.17}_{-0.03}$&$4.8^{+1.3}_{-2.5}\times 10^3$&$1.0_{-0.31}$&1.25~(702)\\

&&&$1.96\pm0.08$&$1.98^{+0.13}_{-0.16}$&$1.17^{+0.13}_{-0.08}$&$2.2^{+1.1}_{-0.93}\times 10^2$&---&1.26~(702)\\

8&761.006--761.079&1612.61&$2.13^{+0.06}_{-0.05}$&$1.78^{+0.06}_{-0.08}$&$0.50\pm0.04$&$1.0^{+0.40}_{-0.30}\times 10^4$&$0.39\pm0.03$&1.02~(694)\\

&&&$2.19\pm0.06$&$1.77^{+0.07}_{-0.08}$&$0.52\pm0.04$&$4.4^{+2.2}_{-1.3}\times 10^3$&---&1.02~(694)\\

9&764.598--764.954&808.25&$1.67\pm0.04$&$2.74^{+0.03}_{-0.40}$&$0.76^{+0.24}_{-0.08}$&$2.1^{+0.9}_{-1.2}\times 10^3$&$0.96^{+0.04}_{-0.50}$&1.02~(568)\\

&&&$1.76^{+0.14}_{-0.12}$&$1.78^{+0.28}_{-0.40}$&$1.21^{+0.11}_{-0.06}$&$2.8\pm1.1\times 10^2$&---&1.03~(568)\\


10&771.600--771.673&1872.09&$1.91^{+0.07}_{-0.05}$&$1.83^{+0.05}_{-0.07}$&$0.51\pm0.04$&$6.1^{+3.0}_{-1.7}\times 10^3$&$0.46^{+0.03}_{-0.04}$&1.15~(691)\\

&&&$2.01\pm0.06$&$1.82^{+0.08}_{-0.07}$&$0.54\pm0.05$&$2.3^{+1.2}_{-0.74}\times 10^3$&---&1.15~(691)\\

11&776.532--776.550&1507.95&$1.92^{+0.03}_{-0.04}$&$2.59^{+0.21}_{-0.33}$&$0.96^{+0.13}_{-0.19}$&$1.7^{+1.8}_{-0.5}\times 10^3$&$0.70^{+0.30}_{-0.39}$&1.07~(683)\\
&&&$2.02\pm0.11$&$1.68^{+0.24}_{-0.40}$&$1.30^{+0.08}_{-0.05}$&$3.9^{+1.3}_{-1.2}\times 10^2$&---&1.07~(683)\\

12&781.013--781.090&1922.34&$2.01^{+0.03}_{-0.02}$&$4.10^{+0.16}_{-0.63}$&$1.10^{+0.19}_{-0.05}$&$1.1^{+0.2}_{-0.4}\times 10^3$&$0.98^{+0.02}_{-0.62}$&1.28~(706)\\
&&&$2.33^{+0.16}_{-0.14}$&$2.49^{+0.45}_{-0.35}$&$1.60^{+0.08}_{-0.07}$&$2.5^{+0.55}_{-0.32}\times 10^2$&---&1.26~(706)\\

13&786.399--786.679&1868.80&$1.74^{+0.03}_{-0.02}$&$2.54^{+0.07}_{-0.06}$&$0.71^{+0.04}_{-0.03}$&$2.0^{+0.4}_{-0.3}\times 10^3$&$0.85^{+0.15}_{-0.31}$&1.07~(670)\\
&&&$1.91\pm0.09$&$1.91^{+0.16}_{-0.21}$&$1.13^{+0.08}_{-0.05}$&$2.1^{+0.79}_{-0.74}\times 10^2$&---&1.08~(670)\\

14&791.866--791.941&1887.39&$1.71^{+0.03}_{-0.04}$&$2.61^{+0.07}_{-0.10}$&$0.66^{+0.14}_{-0.04}$&$1.9^{+0.4}_{-0.9}\times 10^3$&$1.0_{-0.3}$&1.05~(666)\\

&&&$1.91\pm0.09$&$2.05^{+016}_{-0.18}$&$1.20^{+0.13}_{-0.08}$&$1.1^{+0.57}_{-0.46}\times 10^2$&---&1.06~(666)\\

15&796.076--796.088&1044.38&$2.20^{+0.03}_{-0.05}$&$2.37^{+0.07}_{-0.15}$&$0.66^{+0.24}_{-0.08}$&$2.2^{+1.9}_{-1.2}\times 10^3$&$0.93^{+0.07}_{-0.38}$&1.16~(660)\\

&&&$2.33^{+0.14}_{-0.12}$&$1.82^{+0.21}_{-0.26}$&$1.15^{+0.12}_{-0.06}$&$1.9^{+0.85}_{-0.93}\times 10^2$&---&1.16~(660)\\

16\footnotemark[a]&796.329--796.341&1054.55&$1.67\pm0.04$&$3.66^{+0.39}_{-0.68}$&$0.96^{+0.21}_{-0.06}$&$7.4^{+2.0}_{-3.2}\times 10^2$&$1.0_{-0.8}$&0.99~(572)\\

&&&$1.80^{+0.05}_{-0.06}$&$2.00$~(fixed)&$1.38^{+0.04}_{-0.05}$&$1.7^{+0.16}_{-0.14}\times 10^2$&---&1.00~(573)\\

17&801.808--801.884&1988.03&$1.75^{+0.04}_{-0.03}$&$2.83^{+0.10}_{-0.21}$&$0.70^{+0.20}_{-0.03}$&$2.0^{+0.5}_{-1.0}\times 10^3$&$1.0_{-0.44}$&1.08~(641)\\

&&&$1.95\pm0.10$&$2.09^{+0.17}_{-0.21}$&$1.20^{+0.10}_{-0.07}$&$1.8^{+0.77}_{-0.62}\times 10^2$&---&1.08~(641)\\

18&806.147--806.151&286.67&$1.82\pm0.03$&$2.06\pm0.04$&$0.78^{+0.01}_{-0.02}$&$1.2\pm0.01\times 10^3$&$0.45^{+0.55}_{-0.17}$&1.02~(451)\\

&&&$1.92^{+0.19}_{-0.16}$&$1.78^{+0.36}_{-0.55}$&$0.91\pm0.08$&$3.8^{+2.2}_{-1.7}\times 10^2$&---&1.02~(451)\\

19&806.216--806.666&1603.02&$1.68\pm0.03$&$2.37\pm0.09$&$0.72^{+0.10}_{-0.18}$&$1.3^{+0.2.1}_{-0.4}\times 10^3$&$0.64^{+0.36}_{-0.16}$&1.20~(670)\\

&&&$1.80\pm0.08$&$1.89^{+0.16}_{-0.20}$&$1.00^{+0.05}_{-0.04}$&$2.1^{+0.63}_{-0.27}\times 10^2$&---&1.22~(670)\\

20\footnotemark[a]&811.545--811.742&2045.45&$1.73\pm0.03$&$1.77^{+2.07}_{-0.76}$&$1.26^{+0.05}_{-0.16}$&$4.0^{+2.1}_{-0.44}\times 10^2$&$0.12^{+0.46}_{-0.03}$&1.12~(670)\\

&&&$1.84\pm0.04$&$2.00$~(fixed)&$1.32\pm0.03$&$2.8^{+0.16}_{-0.14}\times 10^2$&---&1.13~(671)\\

21\footnotemark[a]&823.538--823.865&1347.05&$1.74^{+0.04}_{-0.03}$&$1.59^{+1.84}_{-0.58}$&$1.24^{+0.06}_{-0.15}$&$2.7^{+1.4}_{-0.32}\times 10^2$&$0.12^{+0.39}_{-0.02}$&0.93~(610)\\

&&&$1.87\pm0.04$&$2.00$~(fixed)&$1.31\pm0.03$&$1.7^{+0.13}_{-0.12}\times 10^2$&---&0.94~(611)\\

22\footnotemark[a]&824.331--824.343&1008.10&$1.67\pm0.03$&$2.00$~(fixed)&$1.24^{+0.05}_{-0.04}$&$2.2^{+0.32}_{-0.30}\times 10^2$&$0.13\pm0.03$&1.02~(558)\\

&&&$1.77\pm0.05$&$2.00$~(fixed)&$1.32\pm0.04$&$1.5^{+0.12}_{-0.11}\times 10^2$&---&1.03~(559)\\

23&826.457--826.526&1922.32&$1.88^{+0.04}_{-0.05}$&$2.53^{+0.05}_{-0.06}$&$0.53^{+0.07}_{-0.04}$&$4.0^{+1.0}_{-2.1}\times 10^3$&$1.0_{-0.29}$&1.11~(669)\\

&&&$2.12^{+0.08}_{-0.09}$&$2.22^{+0.12}_{-0.14}$&$1.05^{+0.15}_{-0.08}$&$1.1^{+0.78}_{-0.60}\times 10^2$&---&1.12~(669)\\


24&831.598--831.608&826.09&$1.90\pm0.05$&$2.78^{+0.32}_{-0.24}$&$0.92^{+0.07}_{-0.22}$&$5.0\pm0.7\times 10^2$&$0.58^{+0.42}_{-0.38}$&1.22~(538)\\

&&&$2.02^{+0.18}_{-0.16}$&$1.88^{+0.34}_{-0.63}$&$1.17^{+0.10}_{-0.06}$&$1.4^{+0.57}_{-0.58}\times 10^2$&---&1.22~(538)\\


25\footnotemark[*]\footnotemark[a]&850.514--850.583&1822.17&$1.84\pm0.03$&$1.24^{+1.35}_{-0.23}$&$1.09^{+0.02}_{-0.06}$&$3.9^{+0.23}_{-0.28}\times 10^2$&$0.08^{+0.07}_{-0.01}$&1.20~(583)\\

&&&$1.94\pm0.03$&$2.00$~(fixed)&$1.09\pm0.02$&$3.1^{+0.22}_{-0.20}\times 10^2$&---&1.21~(584)\\


26\footnotemark[a]&860.656--860.734&2452.92&$1.95\pm0.02$&$2.00$~(fixed)&$1.14\pm0.01$&$4.3^{+0.31}_{-0.25}\times 10^2$&$0.07\pm0.01$&1.18~(678)\\

&&&$2.02\pm0.02$&$2.00$~(fixed)&$1.16\pm0.02$&$3.6^{+0.17}_{-0.16}\times 10^2$&---&1.18~(678)\\


27&870.848--870.864&1326.27&$1.96\pm0.02$&---&$1.21\pm0.01$&$3.1\pm0.13\times 10^2$&---&1.03~(602)\\

28&880.234--880.320&1856.07&$1.98\pm0.02$&---&$1.19\pm0.01$&$3.5\pm0.14\times 10^2$&---&1.04~(601)\\

29&890.241--890.386&2161.76&$1.91\pm0.02$&---&$1.20\pm0.01$&$4.6^{+0.17}_{-0.16}\times 10^2$&---&1.24~(634)\\

30&900.439--900.575&1909.55&$1.93\pm0.02$&---&$1.17\pm0.01$&$3.5^{+0.12}_{-0.11}\times 10^2$&---&1.22~(645)\\









31\footnotemark[*]&910.421--910.961&982.10&$1.88\pm0.02$&---&$1.12\pm0.01$&$4.0^{+0.19}_{-0.18}\times 10^2$&---&1.14~(534)\\

32&920.035--920.038&304.79&$1.92\pm0.05$&---&$1.08\pm0.02$&$4.2^{+0.38}_{-0.35}\times 10^2$&---&1.03~(407)\\

33&920.567--920.571&334.14&$1.92^{+0.05}_{-0.04}$&---&$1.07\pm0.02$&$4.2^{+0.38}_{-0.34}\times 10^2$&---&1.06~(417)\\

34\footnotemark[**]&924.412--924.483&857.09&$2.33\pm0.04$&---&$1.14\pm0.01$&$3.4^{+0.22}_{-0.21}\times 10^2$&---&1.15~(461)\\

35\footnotemark[*]&932.616--932.695&2073.78&$1.86\pm0.02$&---&$1.03\pm0.01$&$3.8\pm0.15\times 10^2$&---&1.21~(522)\\

36&940.339--940.413&1680.56&$2.03\pm0.02$&---&$1.00\pm0.01$&$4.6^{+0.21}_{-0.20}\times 10^2$&---&1.07~(548)\\

37&953.926--954.000&1793.00&$1.92\pm0.02$&---&$0.99\pm0.01$&$3.9^{+0.18}_{-0.17}\times 10^2$&---&1.17~(546)\\

38&954.326--954.398&1751.47&$1.81\pm0.02$&---&$0.99\pm0.01$&$4.0^{+0.18}_{-0.17}\times 10^2$&---&1.23~(542)\\

39&955.524--956.143&3535.17&$2.03\pm0.01$&$4.18\pm0.07$&$0.814^{+0.004}_{-0.003}$&$8.41\pm0.03\times 10^2$&$0.30^{+0.12}_{-0.19}$&1.15~(613)\\

&&&$2.23\pm0.12$&$2.84^{+0.25}_{-0.38}$&$0.90^{+0.02}_{0.01}$&$4.7^{+0.58}_{-0.55}\times 10^2$&---&1.15~(613)\\

40&959.788--959.935&540.31&$1.92\pm0.05$&---&$0.97\pm0.02$&$3.6^{+0.35}_{-0.32}\times 10^2$&---&1.04~(395)\\
41&960.323--961.201&3102.55&$1.72\pm0.02$&---&$0.96\pm0.01$&$3.5\pm0.13 \times 10^2$&---&1.18~(578)\\

42&962.458--962.468&809.93&$1.60\pm0.04$&---&$0.93\pm0.01$&$3.5^{+0.27}_{-0.25}\times 10^2$&---&1.08~(409)\\


43\footnotemark[b]&966.0--994.0&44300.11&$2.00$~(fixed)&$2.00$~(fixed)&$0.72^{+0.09}_{-0.08}$&$6.4^{+5.8}_{-2.9}\times 10^2$&$0.07\pm0.02$&1.24~(50)\\
&&&$2.00$~(fixed)&$2.00$~(fixed)&$0.73^{+0.09}_{-0.08}$&$5.5^{+4.9}_{-2.5}\times 10^2$&---&1.24~(50)\\


\hline
\multicolumn{8}{@{}l@{}}{\hbox to 0pt{\parbox{180mm}{\footnotesize
\par\noindent
\footnotemark[$(0)$]MJD noted in this table = actual MJD - $56000$.
\par\noindent
\footnotemark[$(1)$]$N_\mathrm{H}$: Hydrogen column density.
\par\noindent
\footnotemark[$(2)$]$\Gamma$: Photon index.
\par\noindent
\footnotemark[$(3)$]$T_\mathrm{in}$: Innermost temperature of accretion disk.
\par\noindent
\footnotemark[$(4)$]$Norm\ of\ Diskbb$: Normalization of disk component, given by $(r_{in}/D_{10})^2 \cos\theta$, where $r_{in}$ is ``an apparent'' inner disk radius in km, $D_{10}$ the distance to the source in units of 10 kpc, and $\theta$ the angle of the disk ($\theta
= 0$ is face-on).
\par\noindent
\footnotemark[$(5)$]FracStr: fraction of scattering.
\par\noindent
\footnotemark[$*$] The data in the energy range of 0.5-1.0 keV is excluded when the $\chi^2_{\nu}>1.26$. The change of resultant parameters is negligible after excluding this energy band.
\par\noindent
\footnotemark[$**$] The data in the energy range of 1.8-2.5 keV is excluded when the $\chi^2_{\nu}>1.26$. The change of resultant parameters is negligible after excluding this energy band.
\par\noindent
\footnotemark[$a$] Photon index in the combined model, \texttt{diskbb+power-law}, or the convolved model, \texttt{simpl$\times$diskbb}, is kept fixed, so as to constrain the normalization of power-law component or the fraction of scattering of Comptonized component. Parameters of thermal component are not affected by this action.
\par\noindent
\footnotemark[$b$] Hydrogen column density and photon index are kept fixed, so as to constrain other parameters.
\par
}\hss}}
  \end{longtable}
  \end{center}

 \begin{table}
 \caption {Fitting results of Swift/XRT spectra during high/soft state with \texttt{kerbb} model (spin ``$a$" and inclination ``$i$" are fixed at given values).} \label{tab:appen1} 

  \begin{tabular}{@{\extracolsep{\fill}}ccc|cc|cc|cc}
  \hline
&&&\multicolumn{2}{c}{$a=1\ i=43^{\circ}$}\vline&\multicolumn{2}{c}{$a=1\ i=60^{\circ}$}\vline&\multicolumn{2}{c}{$a=0 \ i=33^{\circ}$}\\
\hline
Data ID&\footnotemark[$(0)$]MJD&Exposure&$M_{1}$&$\chi^2_{\nu}~(dof)$&$M_{2}$&$\chi^2_{\nu}~(dof)$&$M_{3}$&$\chi^2_{\nu}~(dof)$\\
&start--end&(s)&($\mathrm{M_{\odot}}$)&&($\mathrm{M_{\odot}}$)&&($\mathrm{M_{\odot}}$)&\\
\hline
27&870.848--870.864&1326.27&$8.22^{+0.22}_{-0.21}$&1.01(602)&$16.20^{+0.39}_{-0.38}$&1.01(602)&$1.70\pm0.04$&1.00(602)\\

28&880.234--880.320&1856.07&$8.78^{+0.21}_{-0.22}$&1.11(601)&$17.41\pm0.40$&1.06(601)&$1.82\pm0.04$&1.04(601)\\

29&890.241--890.386&2161.76&$10.23\pm0.22$&1.36(634)&$20.12^{+0.41}_{-0.40}$&1.30(634)&$2.10\pm0.04$&1.26(634)\\

30&900.439--900.575&1909.55&$8.88^{+0.18}_{-0.17}$&1.23(645)&$17.42^{+0.33}_{-0.32}$&1.22(645)&$1.82\pm0.03$&1.20(645)\\

32&920.035--920.038&304.79&$10.00^{+0.53}_{-0.50}$&1.02(407)&$19.34^{+0.0.96}_{-0.91}$&1.02(407)&$2.01^{+0.10}_{-0.09}$&1.02(407)\\

33&920.567--920.571&334.14&$10.10^{+0.52}_{-0.50}$&1.05(417)&$19.52^{+0.94}_{-0.89}$&1.05(417)&$2.02^{+0.10}_{-0.08}$&1.05(417)\\

36&940.339--940.413&1680.56&$10.87^{+0.28}_{-0.27}$&0.98(548)&$20.68^{+0.50}_{-0.48}$&1.01(548)&$2.10^{+0.03}_{-0.02}$&1.02(548)\\

37&953.926--954.000&1793.00&$9.95^{+0.27}_{-0.26}$&1.08(546)&$19.00^{+0.48}_{-0.46}$&1.12(546)&$1.97^{+0.04}_{-0.05}$&1.12(546)\\

38&954.326--954.398&1751.47&$10.02\pm0.26$&1.18(542)&$19.17^{+0.47}_{-0.46}$&1.20(542)&$1.98^{+0.05}_{-0.04}$&1.18(542)\\

40&$959.788-959.93$&540.31&$9.67^{+0.53}_{-0.50}$&0.98(395)&$18.39^{+0.94}_{-0.89}$&1.01(395)&$1.90^{+0.10}_{-0.09}$&1.01(395)\\

41&960.323--961.201&3102.55&$9.48^{+0.21}_{-0.20}$&1.14(578)&$18.09^{+0.37}_{-0.36}$&1.16(578)&$1.87\pm0.04$&1.14(578)\\


42&962.458--962.468&809.93&$9.60^{+0.42}_{-0.40}$&1.14(409)&$18.24^{+0.74}_{-0.72}$&1.10(409)&$1.88^{+0.07}_{-0.08}$&1.09(409)\\
\hline


\end{tabular}
\begin{tabnote}
\footnotemark[$(0)$]MJD noted in this table = actual MJD - $56000$.\\
Note: All spectra are fitted with \texttt{tBabs $\times$ kerrbb} model. The hydrogen column density $N_{\mathrm{H}}$ almost centres at $2\times10^{22}\ \mathrm{cm}^{-2}$, except the $N_{\mathrm{H}}$ is $1.64\times10^{22}-1.68\times10^{22}\ \mathrm{cm}^{-2}$ for the last spectrum (data ID: 42).
\end{tabnote}
\end{table}

  \begin {table}
\caption {Best-fit spins of Swift/XRT spectrum (data ID: 27 and 36) in \texttt{kerrbb} model, where black hole mass is given by $L_{\mathrm{max}}=L_{\mathrm{Edd}}$
and inclination is fixed to $60^{\circ}$.} \label{tab:appen2} 

\begin{tabular}{@{\extracolsep{\fill}}cc|cc|cc}
\hline
&&\multicolumn{2}{c}{ID\ 27}\vline&\multicolumn{2}{c}{ID\ 36}\\
\hline
D&M&$a1$&$\chi^2_{\nu}~(dof)$&$a2$&$\chi^2_{\nu}~(dof)$\\
(kpc)&($\mathrm{M_{\odot}}$)&&\\
\hline
6.0&2.0&$0.10^{+0.03}_{-0.04}$&0.99(602)&$-0.34^{+0.05}_{-0.04}$&0.99(548)\\
6.5&2.3&$0.18^{+0.03}_{-0.04}$&0.99(602)&$-0.22\pm0.04$&0.99(548)\\
7.0&2.7&$0.29\pm0.03$&0.99(602)&$-0.07\pm0.04$&0.99(548)\\
7.5&3.1&$0.36^{+0.03}_{-0.02}$&1.00(602)&$0.04\pm0.03$&0.99(548)\\
8.0&3.5&$0.42^{+0.03}_{-0.02}$&1.00(602)&$0.12\pm0.03$&0.98(548)\\
8.5&4.0&$0.50^{+0.02}_{-0.03}$&1.00(602)&$0.22^{+0.02}_{-0.04}$&0.98(548)\\
\hline

\end{tabular}
\begin{tabnote}
Note: The spectra were fitted with \texttt{tBabs $\times$ kerrbb} model. The hydrogen column density $N_{\mathrm{H}}$ falls within $2.04\times10^{22}-2.14\times10^{22}\ \mathrm{cm}^{-2}$.
\end{tabnote}
\end{table}



\begin{thebibliography}{}
\bibitem [Borozdin, K. N., et al. (1998)]{key}
 Borozdin, K. N., Revnivtsev, M. G., \& Trudolyubv, S. P. 1998, Astron Lett, 24, 435
 
 \bibitem [Borozdin, K., & Trudolyubov, S. P. (2000)]{key}
Borozdin, K., \& Trudolyubov, S. P. 2000, ApJL, 533, 431

\bibitem [Burrows, D. N., et al. (2005)]{key}
Burrows, D. N., et al. 2005, Space Sci Rev., 120, 165


\bibitem [Coe, M. J., et al. (1976)]{key}
Coe, M. J., Engel, A. R., \& Quenby, J. J. 1976, Nature, 259: 544-545

\bibitem [Corral-Santana J. M., et al. (2016)]{key}
Corral-Santana J. M., Casares J., Mun?oz-Darias T., Bauer F. E., Mart??nez-Pais I. G., \& Russell D. M. 2016, A\&A, 587, A61

\bibitem [Dennerl K., \&  Greiner J. (1996)]{key}
Dennerl K., Greiner J. 1996, IAUC , 6426, 1

\bibitem [Fabian, A. C., et al. (1989)]{key}
Fabian, A. C., Rees, M. J., Stella, L., \& White, N. E. 1989, MNRAS, 238, 719

\bibitem [Fender, R. P., \& Belloni, T. (2004)]{key}
Fender, R. P., \& Belloni, T. 2004, Annu. Rev. Astron. Astrophys, 42: 317-64

\bibitem [Filippova, E., et al. (2014)]{key}
Filippova, E., et al. 2014, ATel, 5991

\bibitem [Frank, J., et al. (1987)]{key}
Frank, J., King, A. R., \& Lasota, J. P. 1987, A\&A, 178: 137-142

\bibitem [F\"urs, F., et al. (2016)]{key}
F\"urs, F., et al. 2016, ApJ, 832, 115

\bibitem [Greiner, J., et al. (1996)]{key}
Greiner, J., Dennerl, K., \& Predehl, P. 1996, A\&A, 314, 121

\bibitem [Hjellming, R. M., et al. (1996)]{key}
Hjellming, R. M., Rupen, M.P., Marti, J., Mirabel, F., \& Rodriguez, I. F. 1996, IAU Circ., 6383


\bibitem [Kubota, A., et al. (1998)]{key}
Kubota, A., Tanaka, Y., Makishima, K., Ueda, Y., Dotani, T., Inoue, H., \& Yamaoka, K. 1998, PASJ, 50, 667


\bibitem [Krimm, H., et al. (2014)]{key}
Krimm, H., Barthelmy, S. D., Baumgarther, W., et al. 2014, ATel, 5986

\bibitem [Li, LX et al. (2005)]{key}
Li, LX., Zimmerman, E. R., Narayan, R., \& McClintock. J. E. 2005, ApJSuppl, 157, 335.

\bibitem [Maccarone, T. J. (2003)]{key}
Maccarone, T. J. 2003, A\&A, 409, 697

\bibitem [Makishima, et al. (1986)]{key}
Makishima et al. 1986, ApJ, 308, 635

\bibitem [Markert. T. H. (1973)]{key}
Markert. T. H., Canizares, C. R., Clark, G. W., et al. 1973, ApJ, 184, L67

\bibitem [Markert. T. H. (1973)]{key}
Marti, J.,  Mirabel, I. F., Duc, P.-A., \&  Rodriguez, L. F. 1997, A\&A, 323, 158

\bibitem [Matsuoka, M., et al. (2009)]{key}
Matsuoka, M., et al. 2009, PASJ, 61, 999

\bibitem [McClintock, J. E., et al. (2014)]{key}
McClintock, J. E., Narayan, R., \& Steiner, J. F. 2014, Space Sci. Rev., 183, 295

\bibitem [Mereminskiy, I. A., et al. (2017)]{key}
Mereminskiy, I. A., Filippova, E. V., Krivonos, R. A., Grebenev, S. A., Burenin, R. A., \& Sunyaev, R. A. 2017, Astron let, n..3

\bibitem [Mihara, et al. (2011)]{key}
Mihara, T., et al. 2011, PASJ, 63, S623

\bibitem [Miller, J. M., et al. (2015)]{key}
Miller, J. M., et al. 2015, ApJL, 799, L6

\bibitem [Mitsuda, et al. (1984)]{key}
Mitsuda et al. 1984, PASJ, 36, 741


\bibitem [Nakahira, S., et al. (2012)]{key}
Nakahira, S. et al, 2012, PASJ, 64, L13


\bibitem [Paul, J., et al. (1996)]{key}
Paul, J., Bouchet, L., Churazov, E., \& Sunyaev, R. 1996, IAU Circ, 6438


\bibitem [Remillard, R. A., \& McClintock, J. E. (2006)]{key}
Remillard, R. A., \& McClintock, J. E. 2006, A\&A, 44, 49R

\bibitem [Reynolds, C. S. (2014)]{key}
Reynolds, C. S. 2014, Space Sci Rev., 183, 277

\bibitem [Reynolds, M. T., \& Miller, J. M. (2013)]{key}
Reynolds, M. T., \& Miller, J. M. 2013, ApJ, 769, 16R

\bibitem [Shimura T. \& Takahara F. (1995)]{key}
Shimura T., Takahara F. 1995, ApJ , 445, 780

\bibitem [Sobczak, G. J., et al (2000)]{key}
Sobczak, G. J., McClintock, J. E., Remillard, R. A., et al. 2000, ApJ, 531: 537-545

\bibitem [Steiner, J. F., et al. (2009)]{key}
Steiner, J. F., Narayan, R., McClintock, J. E. \& Ebisaua, K. 2009, PASP, 121, 279

\bibitem [Steiner, J. F., et al. (2016)]{key}
Steiner J.-F., Remillard R.-A., García J.-A., McClintock J.-E. 2016, ApJ , 829, L22

\bibitem [Tananbaum. H., et al. (1972)]{key}
Tananbaum. H., et al. 1972, ApJL, 177, L5

\bibitem [Tsunemi, et al. (1989)]{key}
Tsunemi, H., Kitamoto, S., Okamura, S., \& Diane, R. 1989, ApJ, 337L, 81T.


\bibitem [Tsunemi, et al. (2010)]{key}
Tsunemi, H., Tomida, H., Katayama, H., Kimura, M., Daikyuji, A., Miyaguchi, K., Maeda, K., \& the MAXI team. 2010, PASJ, 62, 1371

\bibitem [Vargas, M., et al. (1997)]{key}
Vargas, M., Goldwurm, A., Laurent, P., et al. 1997, ApJL, 476, L23

\bibitem [Wijnands, R., et al. (2001)]{key}
Wijnands, R., Mendez, M., Miller, J. M., \& Homan, J. 2001, MNRAS, 328, 451

\bibitem [Zhang, S. N., et al. (1997)]{key}
Zhang, S. N., Cui, W. \& Chen, W. 1997, ApJ, 492: L155-L158


\end{thebibliography}
\end{document}